\documentclass[aps,prd,twocolumn,showpacs,floatfix,preprintnumbers,amsfont,amsmath,amssymb,nofootinbib,superscriptaddress]{revtex4-2}%

\usepackage{aas_macros}
\usepackage{graphicx}
\usepackage{dcolumn}
\usepackage{bm}
\usepackage[normalem]{ulem}
\usepackage[dvipsnames,svgnames,table]{xcolor} 

\usepackage{xcolor}

\definecolor{colorLink}{rgb}{0.9,0,0} %
\definecolor{colorCite}{rgb}{0,0.7,0} %
\definecolor{colorURL} {rgb}{0,0,0.8} %
\usepackage[colorlinks=true,linktocpage=true,linkcolor=colorLink,citecolor=colorCite,urlcolor=colorURL]{hyperref}
\usepackage{bbold}
\usepackage{multirow}
\usepackage{fontawesome}
\usepackage{algorithm2e}
\usepackage{algpseudocode}
\usepackage{orcidlink}
\newcommand{\be}{\begin{equation}}
\newcommand{\ee}{\end{equation}}

\newcommand{\sk}[1]{}

\DeclareMathSymbol{\mhyphen}{\mathord}{AMSa}{"39}

\graphicspath{{./}{figs/}}

\begin{document}

\title{Improving low-latency multi-messenger follow-up of neutron star--black hole mergers with mode-by-mode filtering}

\author{Francesco Iacovelli\texorpdfstring{\,}{ }\orcidlink{0000-0002-4875-5862}}
\email{fiacovelli@jhu.edu}
\affiliation{\mbox{Department of Physics and Astronomy, Johns Hopkins University,
3400 N. Charles Street, Baltimore, Maryland, 21218, USA}}

\author{Digvijay Wadekar\texorpdfstring{\,}{ }\orcidlink{0000-0002-2544-7533}}
\affiliation{\mbox{Center for Gravitational Physics, University of Texas at Austin, Austin, TX 78712, USA}}
\affiliation{\mbox{Department of Physics and Astronomy, Johns Hopkins University,
3400 N. Charles Street, Baltimore, Maryland, 21218, USA}}

 \author{Javier Roulet\texorpdfstring{\,}{ }\orcidlink{0000-0003-3268-4796}}
\affiliation{\mbox{Kavli Institute for Cosmological Physics, The University of Chicago, 5640 South Ellis Avenue, Chicago, IL 60637, USA}}
\affiliation{\mbox{School of Natural Sciences, Institute for Advanced Study, 1 Einstein Drive, Princeton, NJ 08540, USA}}

\author{Emanuele Berti\texorpdfstring{\,}{ }\orcidlink{0000-0003-0751-5130}}
\affiliation{\mbox{Department of Physics and Astronomy, Johns Hopkins University,
3400 N. Charles Street, Baltimore, Maryland, 21218, USA}}

\author{Alessandra Corsi\texorpdfstring{\,}{ }\orcidlink{0000-0001-8104-3536}}
\affiliation{\mbox{Department of Physics and Astronomy, Johns Hopkins University,
3400 N. Charles Street, Baltimore, Maryland, 21218, USA}}

 \date{\today}

\begin{abstract}
Rapid parameter estimation for neutron star-black hole (NSBH) mergers is essential for deciding whether, where, and how electromagnetic facilities should follow up gravitational-wave alerts. Current low-latency analyses typically use only the dominant quadrupole harmonic, leaving strong degeneracies among luminosity distance, inclination, and intrinsic binary parameters. We show that mode-by-mode filtering of the $(2,2)$, $(3,3)$, and $(4,4)$ signal-to-noise-ratio (SNR) time series enables low-latency marginalization over higher-order-mode information at a computational cost comparable to quadrupole-only analyses\footnote{A public implementation and example workflow are available at \url{https://github.com/JayWadekar/flywheel}.}. Applied to simulated NSBH detections in a LIGO-Virgo network at design sensitivity, our method improves constraints on luminosity distance, viewing angle, localization volume, and source-frame secondary mass, thereby sharpening crucial estimates of electromagnetic detectability and host-galaxy association. We also validate the approach on public data for previously detected NSBH events, finding the largest improvement for the asymmetric, higher-SNR event GW190814.
\href{https://github.com/JayWadekar/flywheel}{\faGithub}
\end{abstract}
\maketitle

\emph{Introduction---} Gravitational waves (GWs) from compact binary coalescences (CBCs) have transformed astronomy since their first detection in 2015~\cite{Abbott:2016blz}. The LIGO-Virgo-KAGRA (LVK)~\cite{LIGOScientific:2014pky,Virgo:2014yos,Aso:2013eba} Collaboration GW transient catalog now contains over 390 candidates~\cite{LIGOScientific:2025slb,LIGOScientific:2026wfs}, predominantly binary black holes. Binary neutron star (BNS) mergers remain rarer, with only two confident detections~\cite{LIGOScientific:2017vwq,LIGOScientific:2020aai}. The joint observation of GW170817 with electromagnetic (EM) emission~\cite{LIGOScientific:2017ync} marked the beginning of the era of multi-messenger GW astronomy, demonstrating the power of combined observations to constrain nuclear physics~\cite{LIGOScientific:2017vwq,LIGOScientific:2018cki}, cosmology~\cite{LIGOScientific:2017adf}, and relativistic astrophysics.

Neutron star-black hole (NSBH) binaries are a promising, yet elusive, multi-messenger target. Three likely NSBH candidates have been reported~\cite{LIGOScientific:2021qlt,KAGRA:2021vkt,LIGOScientific:2024elc} along with a few marginal ones~\cite{Olsen:2022pin,LIGOScientific:2021usb,KAGRA:2021vkt}, but no confirmed EM counterpart has been identified. Multi-messenger observations of NSBHs, albeit rare with current detectors~\cite{Colombo:2023une}, could constrain the cosmic expansion history via standard-siren measurements~\cite{Vitale:2018wlg}, probe supranuclear-density matter through tidal disruption~\cite{Clarke:2023rrm}, and shed light on the progenitor channels linking stellar evolution to compact remnants~\cite{Broekgaarden:2021efa}. Realizing this potential requires rapid, accurate parameter estimation (PE) to guide time-sensitive EM follow-up by facilities such as the Neil Gehrels \emph{Swift} Observatory~\cite{SwiftScience:2004ykd}, ULTRASAT~\cite{Shvartzvald:2023ofi}, SVOM~\cite{Atteia:2021izu}, and the Einstein Probe~\cite{EinsteinProbeTeam:2015bcj}, as well as wide-field optical searches with the Zwicky Transient Facility~\cite{Anand:2020eyg}, the Vera C. Rubin Observatory~\cite{Andreoni:2021epw}, GOTO~\cite{Gompertz:2020cur}, DECam/GROWTH~\cite{Goldstein:2019roe}, and GRANDMA~\cite{Antier:2020nuy}.

Current low-latency LVK searches and alert products provide sky localization, distance estimates, and source classification within seconds to minutes~\cite{Chaudhary:2023vec,Sachdev:2019vvd,Nitz:2018rgo,Adams:2015ulm,Chu:2020pjv,Villa-Ortega:2022qdo,LIGOScientific:2026ifv}, with rapid localizations commonly produced by the BAYESian TriAngulation and Rapid localization (\href{https://git.ligo.org/lscsoft/ligo.skymap}{\texttt{BAYESTAR}}) pipeline~\cite{Singer:2015ema,Singer:2016eax,Singer:2016erz} (see also Ref.~\cite{Essick:2014wwa}). These analyses model only the dominant $(\ell,|m|)=(2,2)$ GW harmonic. For asymmetric binaries like NSBHs, where mass ratios $q \equiv m_1/m_2$ can exceed $\sim\!5$--$10$, subdominant harmonics %
carry useful information~\cite{Capano:2013raa,Mills:2020thr,Fairhurst:2023idl,Wadekar:2023gea,Yi:2025pxe,Yi:2026ucv,Mehta:2025jiq,Mehta:2025oge,Wadekar:2023kym,Chandra:2022ixv,CalderonBustillo:2022ldv,Sinha:2025vmc}. These modes depend differently on the source parameters and can therefore break degeneracies. In quadrupole-only waveforms, luminosity distance ($d_L$) correlates strongly with inclination ($\iota$), while mass ratio correlates with effective spin ($\chi_{\rm eff}$)~\cite{Cutler:1994ys,Usman:2018imj,CalderonBustillo:2020kcg}. These degeneracies broaden localization volumes, weaken constraints on the nature of the secondary object (NS vs. BH), and degrade inclination estimates, all of which affect follow-up decisions.

Higher-order modes (HM) can break these degeneracies even in low-latency analyses. Mode-by-mode matched filtering~\cite{Wadekar:2024zdq} isolates the contributions of individual harmonics, enabling rapid inference when combined with efficient marginalization techniques~\cite{Roulet:2024hwz}. For NSBHs, improved constraints on $d_L$ and $\iota$ inform crucial observational strategies/choices to: ({i}) efficiently identify the host from galaxy catalogs~\cite{Gehrels:2015uga}; ({ii}) determine if viewing angles favor detectable $\gamma$-ray/X-ray/radio emission from jets, and how long and deep to observe~\cite{Sarin:2022cmu,Salafia:2022dkz,Morsony:2023afu}; ({iii}) assess if the system lies within realistic EM search horizons for kilonov\ae{} (KNe)~\cite{Carracedo:2020xhd,Chase:2021ood,Andreoni:2021jis,Andreoni:2023xlv,Ahumada:2024qpr,Stevenson:2025fqt}. Additionally, tighter mass-ratio constraints help source classification and inference on the potential tidal disruption of the secondary (affecting KN brightness)~\cite{Foucart:2018rjc,Barbieri:2019kli,Gompertz:2023kjl}, reducing false alarms for resource-intensive observing campaigns.

In this work, we show that including HMs in rapid PE of NSBH mergers yields substantially improved localizations and secondary-mass measurements at a computational cost comparable to existing pipelines. We demonstrate this using simulated detections from the best-fit LVK population model~\cite{LIGOScientific:2024elc} in mock data for the upcoming fifth observing run O5, and later validate the method on real LVK events.

\begin{figure}
    \includegraphics[width=\linewidth]{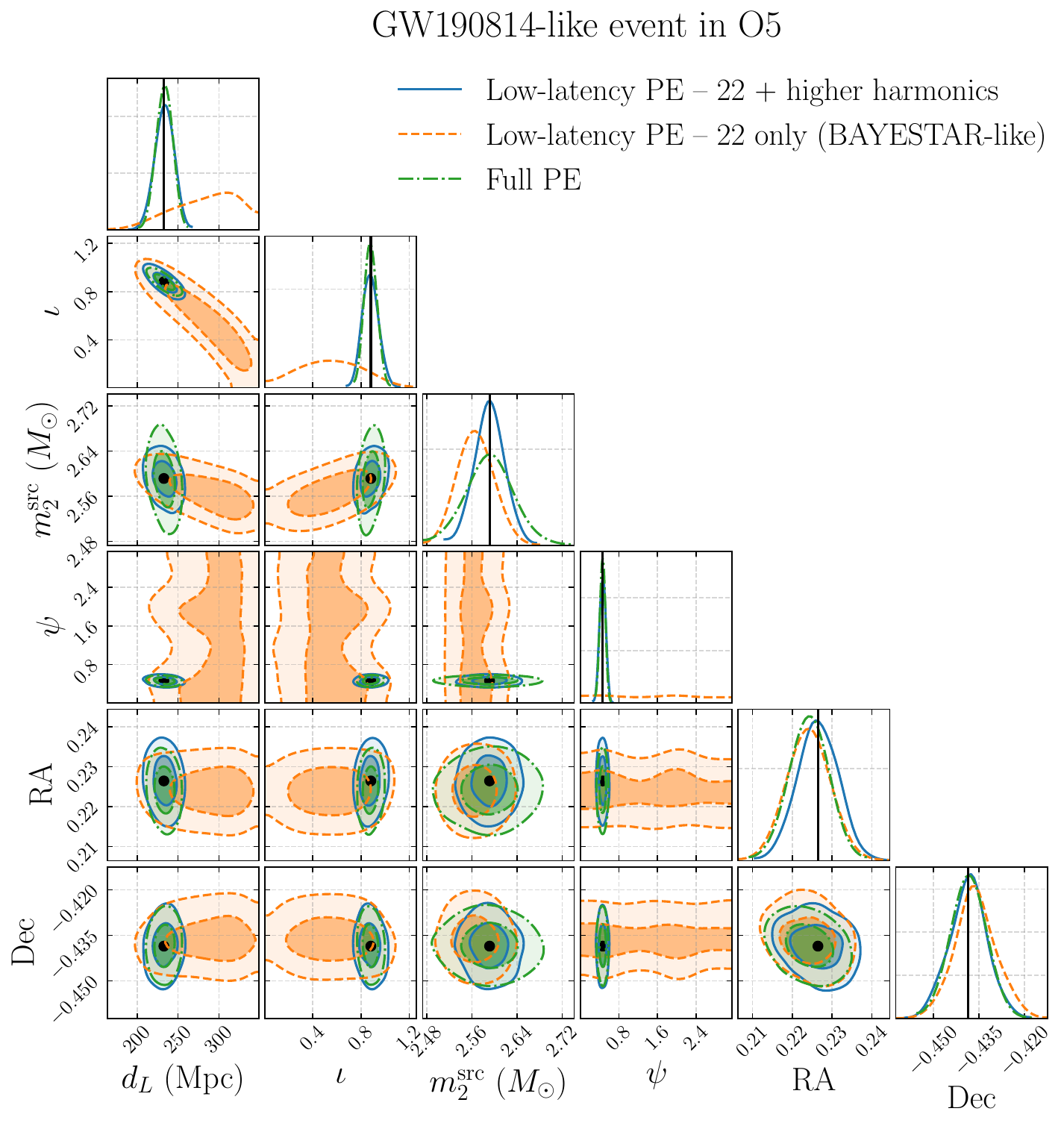}
    \caption{Low-latency PE for a GW190814-like injected signal in a LIGO--Virgo network at O5 design sensitivity. We compare our method including higher harmonics with a quadrupole-only analysis similar to those currently used in low-latency LVK pipelines. We also show full PE results from \href{https://github.com/bilby-dev/bilby}{\texttt{bilby}}. Including HMs recovers the full-PE luminosity-distance and inclination posteriors much more faithfully, while also improving polarization and mildly improving $m_2^{\rm src}$.}
    \label{fig:GW190814}
\end{figure}

\emph{Catalog simulation and SNR time series---} To establish the detectability and characterization prospects for NSBH mergers, we simulate mock populations compatible with the latest LVK results. We employ the \textsc{NSBH-pop} model~\cite{Biscoveanu:2022iue} for masses and aligned-spin components, drawing 100 realizations of 1-year populations out to $z\sim10$ using posterior samples from Ref.~\cite{LIGOScientific:2024elc}. We extrapolate the population to high redshift using a Madau-Dickinson merger-rate density profile with typical parameters~\cite{Madau:2014bja,Madau:2016jbv}. The catalogs contain $\sim\!2.1\times10^4$ to $\sim\!3.4\times10^5$ events each (totaling $\sim\!2.5\times\!10^7$ binaries), ensuring sufficient statistics. Extrinsic parameters are sampled uniformly.
For each event, we compute the signal-to-noise ratio (SNR) in a three-detector network (two LIGO detectors and Virgo) with O5 sensitivities~\cite{KAGRA:2013rdx} using \href{https://github.com/CosmoStatGW/gwfast}{\texttt{gwfast}}~\cite{Iacovelli:2022mbg}. We employ the \textsc{IMRPhenomXHM} waveform~\cite{Garcia-Quiros:2020qpx} (as implemented in the LIGO Algorithm Library, \href{https://git.ligo.org/lscsoft/lalsuite/}{\texttt{LAL}}~\cite{lalsuite}) including the $(\ell,\,|m|)=(2,2)$, $(3,3)$, and $(4,4)$ modes. For systems with ${\rm SNR}\geq8$ ($\sim1.6\times10^4$ in total, redshifts out to $z\sim0.64$, SNRs $\sim\!8$--$85$), we compute zero-noise SNR~\cite{zeronoise_explain} time series for each mode using the \texttt{SNR\_timeseries} module from our publicly available codebase; see Supplemental Material.
While simplified, this zero-noise, perfect-template analysis establishes a clear performance benchmark for our methodology.

\emph{Low-latency marginalized parameter estimation---} Full stochastic-sampling PE (e.g. with \href{https://github.com/bilby-dev/bilby}{\texttt{bilby}~\cite{Ashton:2018jfp}}), which explores the joint posterior over $\sim\!15$ binary parameters, can take hours to days when higher harmonics are included, which is incompatible with low-latency follow-up. Complementary machine-learning approaches such as \href{https://github.com/dingo-gw/dingo}{\texttt{DINGO}}~\cite{Dax:2021tsq,Dax:2024mcn} or \href{https://github.com/jroulet/labrador}{\texttt{labrador}} \cite{Roulet:2026mzz} can also provide rapid compact-binary inference for binary neutron stars, however these do not account for higher modes yet. Our approach instead marginalizes analytically or semi-analytically over parameters that enter the likelihood in simple ways, such as $d_L$ and reference phase, and uses adaptive importance sampling for the remaining extrinsic parameters. This brings the runtime to seconds while retaining faithful posteriors for the parameters most relevant to EM follow-up, as verified against full PE in Fig.~\ref{fig:GW190814} (discussed below). We use \href{https://github.com/JayWadekar/flywheel}{\texttt{flywheel}}, a low-latency version of the \href{https://github.com/jroulet/cogwheel}{\texttt{cogwheel}} library~\cite{Roulet:2024hwz,Roulet:2022kot,Islam:2022afg} previously adapted for the coherent candidate-ranking statistic in the \href{https://github.com/JayWadekar/gwIAS-HM}{\texttt{IAS-HM}} search pipeline~\cite{Wadekar:2024zdq}. The analysis includes the higher-order mode contributions $(\ell, |m|)=(3,3)$ and $(4, 4)$ alongside the dominant $(2, 2)$ harmonic.

The algorithm takes three inputs: (i) $\langle d|h_{\ell m}\rangle$, the SNR time series for each mode template; (ii) $\langle h_{\ell m} | h_{\ell' m'}\rangle$, the mode covariance matrix; and (iii) prior samples for the mode SNR ratios under an assumed astrophysical distribution,
\begin{equation}
R_{\ell m} \equiv \frac{\langle h_{\ell m}(f)|h_{\ell m}(f)\rangle^{1/2}}{\langle h_{22}(f)|h_{22}(f)\rangle^{1/2}}\;.
\label{eq:Rlm}
\end{equation}

Each $(2,2)$ template in a search pipeline corresponds to a region of intrinsic parameter space [$m_1, m_2, \chi_{1z}, \chi_{2z}$]. We use the $R_{\ell m}$ values associated with that region in the marginalization. Details of the sample library construction are given in Supplemental Material.%

We summarize the likelihood and coherent marginalization used in this work in Supplemental Material Algorithm~\ref{app:coherent_score}
(see also Sec.~3A of Ref.~\cite{Wadekar:2024zdq}). We marginalize over the extrinsic binary parameters and the subset of intrinsic parameters which affect the mode amplitude ratios in Eq.~\eqref{eq:Rlm}.
The luminosity-distance integral is evaluated by interpolating a precomputed table~\cite{Singer:2015ema}, the reference-phase integral over $\phi_0$ by trapezoid quadrature, and the remaining integrals over inclination, sky location, and polarization by adaptive importance sampling; see Ref.~\cite{Roulet:2024hwz} for further details on the marginalization procedure.

Once the marginalization integral is computed, posterior samples are obtained by importance resampling the quasi-Monte Carlo samples used in the integral; an illustrative example is given in Supplemental Material, Sec.~\ref{sec:marginalization_sample_example}. 
For ten representative O5 catalog events, one higher-mode marginalization takes $0.13\pm0.03$\,s on a \emph{single} Intel Xeon Gold 6426Y core, and drawing 1000 posterior samples takes an additional $0.11$\,s.

\emph{Inference of catalog properties---} We compare analyses using SNR time series for all three modes against a $(2,2)$-only analysis, mirroring current LVK pipelines such as \texttt{BAYESTAR}.
Figure~\ref{fig:GW190814} compares the two approaches for an event with the same characteristics as GW190814~\cite{LIGOScientific:2020zkf} in an O5 network. GW190814 is one of the most asymmetric mergers observed. We simulate the SNR time series using the maximum-likelihood parameters from public LVK data at our network configuration (optimal matched-filter SNR $\sim\!78$). While sky localization is minimally affected, the $d_L$ and $\iota$ reconstructions improve substantially with HMs, reducing the comoving volume by a factor $\sim\!4.2$. The $(2,2)$-only pipeline's strong $d_L$--$\iota$ correlation exacerbated by the standard PE priors~\cite{pepriors_explain} biases the posterior toward large distances, while including HMs favors smaller distances near the true value.
The HM analysis also increases the log-likelihood ($\ln{\cal L}$) of the drawn samples. Comparing our low-latency PE with a full Bayesian analysis for this simulated event obtained with \href{https://github.com/bilby-dev/bilby}{\texttt{bilby}}~\cite{Ashton:2018jfp} and the nested sampler \href{https://github.com/joshspeagle/dynesty}{\texttt{dynesty}}~\cite{2020MNRAS.493.3132S}, we find that including HMs gives remarkably similar results for the luminosity distance and inclination, while maintaining a low computational cost. Both low-latency analyses shown slightly overconstrain $m_2^{\rm src}$ because they condition on the detector frame intrinsic parameters provided by the search pipeline rather than marginalizing over them; see Supplemental Material for further discussion.%

\begin{figure}[tb]
    \centering
    \includegraphics[width=\linewidth]{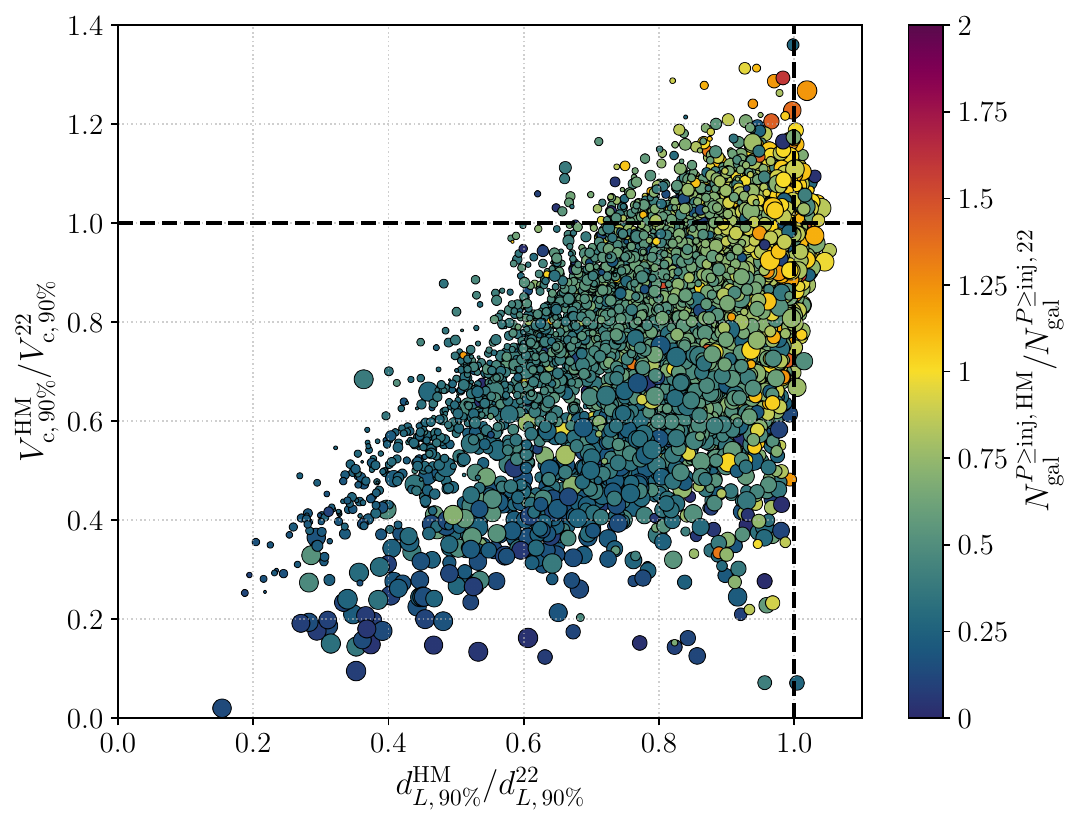}
    \caption{Accurate distance and volume estimates determine how efficiently observers can search for EM counterparts or host galaxies. We show the HM-to-$(2,2)$ ratio of the 90\% c.l. width of the $d_L$ posterior and the 90\% c.l. localization comoving volume for all analyzed NSBH systems; color gives the corresponding ratio of candidate galaxies followed up before reaching the true host (smaller is better), and point size scales with viewing angle. Including HMs usually shrinks the distance--volume search space and reduces the host-galaxy follow-up burden.}
    \label{fig:scatter_dLVc}
\end{figure}

Figure~\ref{fig:scatter_dLVc} shows population-level improvements: for each event, we plot the ratio of 90\% credible-level (c.l.) widths for $d_L$ and comoving volume $V_{\rm c}$ between the two methods. To ensure numerical stability, we estimate $V_{\rm c}$ via $k$-nearest neighbors in a coordinate system designed to minimize correlations.
We use $(d_L, \theta_{\rm net}, \phi_{\rm net})$ coordinates~\cite{Roulet:2022kot}, scaled by directional distance sensitivity. Adding HMs improves constraints for most signals, often by a factor of $\gtrsim\!2$. This is important for EM follow-up because it improves assessments of detectability, reduces the area and volume to search for counterparts, and increases the spatial association probability for the true host galaxy. The point size in Fig.~\ref{fig:scatter_dLVc} is proportional to the viewing angle, $\Theta={\rm min}(\iota,\, 180^\circ-\iota)$.

Some events anomalously show larger $V_{\rm c}$ when including HMs despite narrower $d_L$ posteriors. This is because HMs can narrow the posterior towards larger $d_L$, where the comoving volume is larger; see Supplemental Material. A smaller volume is not necessarily better if it is biased. We therefore consider the number of candidate galaxies an observer would have to follow up before reaching the true host, $N_\mathrm{gal}$. The point colors in Fig.~\ref{fig:scatter_dLVc} show a proxy for $N_\mathrm{gal}$, computed from the number of galaxies in posterior regions with probability density higher than the injection. We use a survey-agnostic convention, assuming a galaxy catalog is already available for the follow-up: galaxies are followed up in order of posterior probability at their location, and their number density is uniform in comoving volume. %

\begin{figure}[tb]
    \centering
    \includegraphics[width=\linewidth]{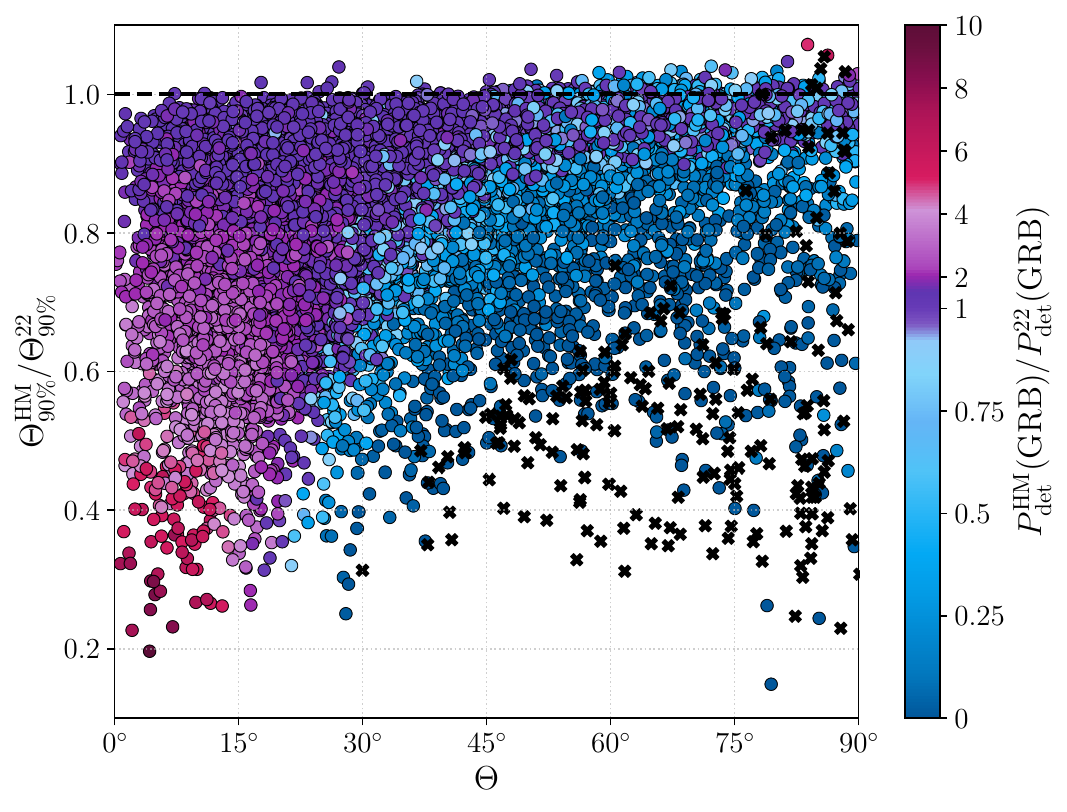}
    \caption{The viewing/inclination angle $\Theta$ controls whether a relativistic jet is likely to be visible, so rapid constraints on $\Theta$ directly affect GRB follow-up decisions. We show the injected $\Theta$ and the HM-to-$(2,2)$ ratio of the posterior width on $\Theta$; color gives the ratio of the {\it Swift}-BAT GRB-detection probability from Eq.~\eqref{eq:f_grb}, and black crosses mark events with vanishing probability in both analyses. We see that HMs sharpen viewing-angle inference and can correctly strengthen or weaken the case for rapid GRB follow-up, depending on the reconstructed orientation.}
    \label{fig:scatter_iota}
\end{figure}

Figure~\ref{fig:scatter_iota} shows that HMs improve viewing-angle constraints across the population, by up to $\sim\!5$ for face-on/off systems, by breaking the $d_L$--$\iota$ degeneracy. This has direct consequences for searches for prompt $\gamma$-ray burst (GRB) or afterglow emission: a rapid reconstruction of $\Theta$ can help determine whether a telescope should slew to the event. As a concrete example, the {\it Swift} Burst Alert Telescope (BAT)~\cite{Barthelmy:2005hs} has a field of view covering $\sim\!10\%$ of the sky, and its slewing time between two sky positions separated by $\theta_{{\rm sep}}$ degrees is $t_{\rm slew} = (25 +2\,\theta_{{\rm sep}})\,{\rm s}$~\cite{Eyles-Ferris:2024bkl}.

Point colors show the ratio of GRB detection probabilities [$P_{\rm det}({\rm GRB})$] between pipelines with and without HMs, computed assuming a universal Gaussian jet (see, e.g., Refs.~\cite{Farah:2019tue,Salafia:2023sjx}):
\begin{equation}
\label{eq:f_grb}
F_{\rm GRB} = \dfrac{A_0}{4\pi d_L^2} \,e^{-\Theta^2/(2\Theta_c^2)}\;,
\end{equation}
with $A_0=2.7\times10^{51}\,{\rm erg\,s}^{-1}$ and $\Theta_c = 3.27^\circ$ obtained from a multimessenger analysis of GW170817~\cite{Troja:2018ruz}. We then impose a detectability cut of $F_{\rm GRB,\,th}=10^{-8}\,{\rm erg\,cm}^{-2}\,{\rm s}^{-1}$, appropriate for {\it Swift}-BAT~\cite{Barthelmy:2005hs}, to calculate $P_{\rm det}({\rm GRB})$.

\begin{figure}[tb]
    \centering
    \includegraphics[width=\linewidth]{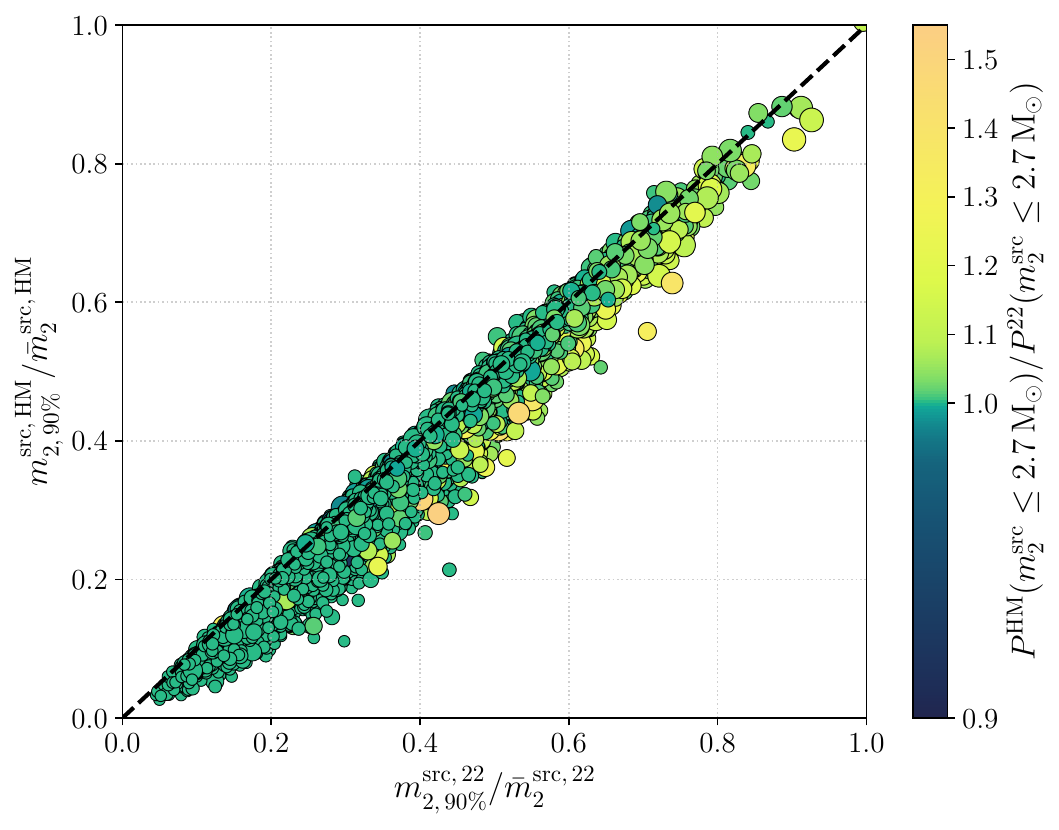}
    \caption{The source-frame secondary mass helps determine whether the lighter object is plausibly a NS and whether tidal disruption could power a kilonova. We compare relative uncertainties on $m_2^{\rm src}$ with HMs ($y$-axis) and without HMs ($x$-axis); color gives the ratio of $P(m_2^{\rm src}<2.7\,{\rm M}_\odot)$ (larger is more accurate), and point size scales with the injected $m_2^{\rm src}$. We see that HMs give modest but useful gains in secondary-mass classification, especially for heavier $m_2^{\rm src}$.}
    \label{fig:scatter_m2relerr_O5_allevents}
\end{figure}

Figure~\ref{fig:scatter_m2relerr_O5_allevents} compares the relative uncertainty in the low-latency secondary-mass constraint, $m_{2,90\%}/\bar{m}_2$. The relative uncertainty ranges from $0.03$ to $0.9$, with median $0.37$.
The quadrupole-only and HM analyses yield similar detector-frame $m_2^\mathrm{det}$ posteriors, so the improvement in $m_2^{\rm src}$ is primarily due to the narrower $d_L$ posterior in the HM case. We assess NS identification probability using a BH/NS mass threshold at $2.7\,{\rm M}_{\odot}$. While most events show modest improvement, higher-mass systems (larger points) gain up to $50\%$ in $P(m_2^{\rm src}\leq2.7\,{\rm M}_{\odot})$ when HMs are included. This can better inform KN follow-up strategies.

Moreover, as we can appreciate from Fig.~\ref{fig:GW190814}, the inclusion of HMs improves the polarization angle $\psi$ constraints by breaking its degeneracy with the orbital phase. This could be relevant for astronomers performing GRB polarization measurements (e.g., the LEAP~\cite{McConnellLEAP} and POLAR~\cite{DeAngelis:2025yyu} missions), which can constrain the emission mechanism if coupled with GW observations~\cite{Kole:2022cdn}.

Finally, not all the simulated NSBH sources could power an EM counterpart. Using the \href{https://git.ligo.org/emfollow/em-properties/em-bright}{\texttt{ligo.em-bright}} pipeline~\cite{Chatterjee:2019avs}, we have verified that $\sim\!34\%$ of the detected systems in our simulation have a probability larger than 10\% to be EM bright. Restricting to these systems, we still find $d_{L,\,90\%}$, $V_{{\rm c},\,90\%}$, and $\Theta_{90\%}$ are reduced by $\sim\!8\%$ on average thanks to HMs, with posteriors shrinking by as much as 70\% for some events.  

\begin{figure}[tb]
    \includegraphics[width=\linewidth]{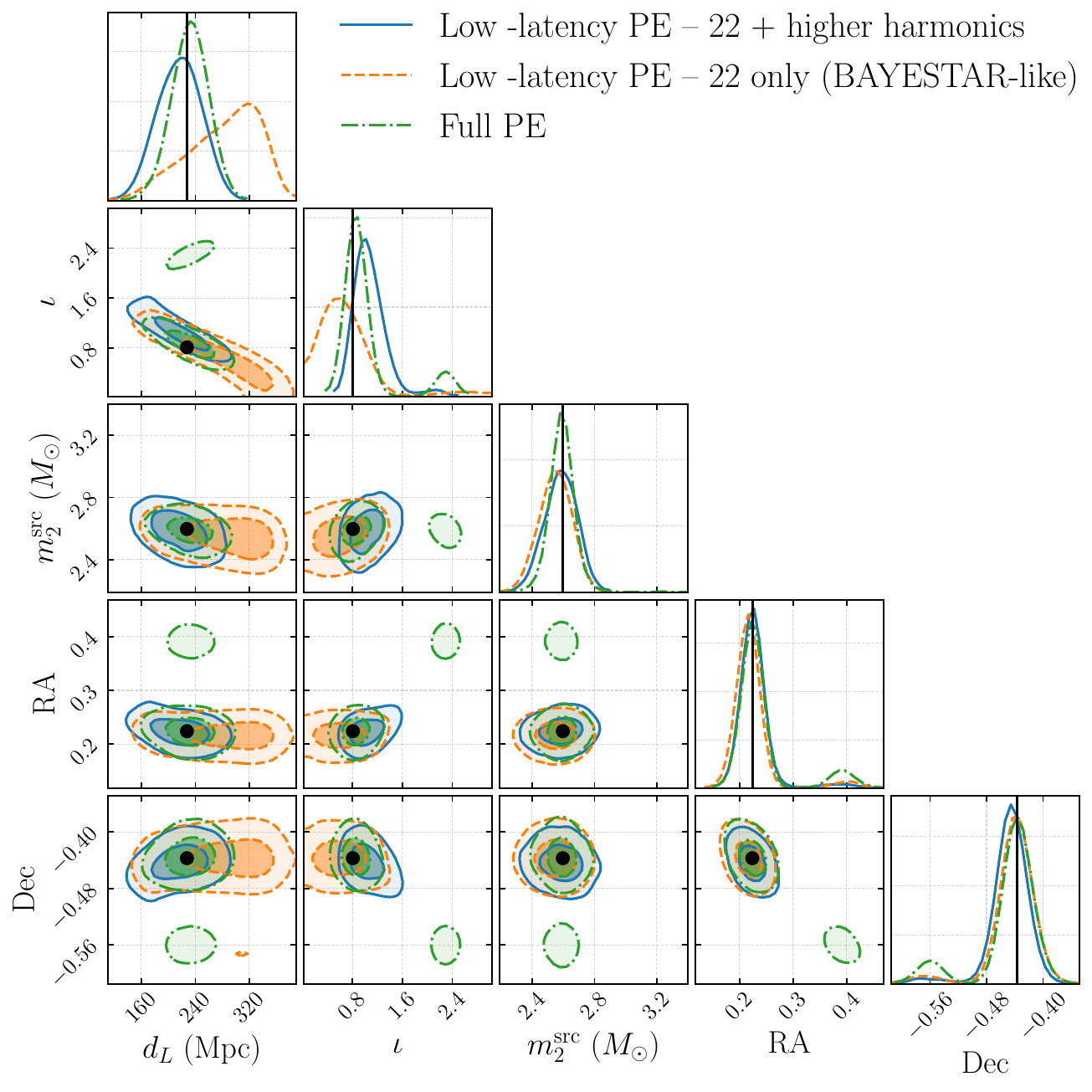}
    \caption{We use our method on the noisy LIGO--Virgo strain data of GW190814 as a real-data test of our method. 
    The black vertical lines show the full-PE posterior sample with the highest matched-filter SNR. We see that HMs move the low-latency posterior toward the full-PE preference for smaller distance and better constrained inclination.}
    \label{fig:corner_GW190814_realdata}
\end{figure}

\emph{Real event analysis---} We apply our pipeline to public LVK data for GW190814~\cite{LIGOScientific:2020zkf} and for the three confident NSBH mergers, GW200105\_162426, GW200115\_042309~\cite{LIGOScientific:2021qlt}, GW230529\_181500~\cite{LIGOScientific:2024elc}. Using \href{https://github.com/gwastro/pycbc}{\texttt{PyCBC}}~\cite{pycbc}, we compute SNR time series for all three modes in windows of 4\,s around each trigger for each active detector, with noise power spectral densities (PSDs) estimated via the Welch method over 32\,s segments preceding the trigger. Mode templates are simulated using the \textsc{IMRPhenomXHM} approximant evaluated at the maximum matched-filter SNR parameters from public posterior samples.

We find that the three NSBH events show no significant improvement with HMs. This is expected given their modest SNRs ($\sim\!11$--$14$) and modest masses ($M_{\rm tot}\lesssim13\,{\rm M}_{\odot}$).
However, GW190814 differs markedly: its higher SNR ($\sim\!25$), larger mass ($M_{\rm tot}\lesssim27.4\,{\rm M}_{\odot}$), and high mass ratio ($q\sim9$) produce a clear $\ell=|m|=3$ spike (${\rm SNR}\sim6$) in the SNR time series.
We show the PE results in Fig.~\ref{fig:corner_GW190814_realdata} and find that using HMs narrows the $d_L$ posterior by a factor of $1.6$ (favoring smaller distances, consistent with full parameter estimation); narrows $\iota$ by a factor of $2$; and reduces the localization comoving volume by a factor of $2.1$. %
This improvement is smaller than for the O5-sensitivity injection of Fig.~\ref{fig:GW190814}, but still substantial.

\emph{Conclusions and future work---} Including HMs in rapid GW parameter estimation improves NSBH binary characterization for EM follow-up. We have developed a ready-to-use pipeline that jointly analyzes the SNR time series of the $(2,2)$, $(3,3)$, and $(4,4)$ harmonics. Applied to simulated O5-sensitivity detections, our pipeline improves the accuracy of luminosity-distance, inclination, and secondary-mass estimation. Crucially, the computational cost is of order a second, comparable to existing $(2,2)$-only methods.

These improvements can be important for a number of EM follow-up strategies. Tighter $d_L$, $\iota$, and $m_2^{\rm src}$ reconstructions enable observers to ({i}) reduce search volumes in galaxy-targeted strategies, potentially decisive for time-limited campaigns; ({ii}) more confidently classify events as NSBH, triggering follow-up protocols; ({iii}) assess EM detectability before using their resources.

We validate our method using data from previously detected NSBH candidates. For GW190814, the low-latency posterior more closely reproduces the full Bayesian posterior, demonstrating the real-data viability of the method. The other three lower-SNR NSBH events show no improvement, consistent with their weaker subdominant mode content. Since our method only requires mode-by-mode SNR time series and their covariance, it can in principle be applied to outputs from any search pipeline that provides these quantities. Our findings motivate generalizing the quadrupole-only low-latency searches currently in use with our HM-based method in the upcoming O5 run.

Several extensions remain for future work. One advantage of HMs is that they enter the detector band earlier than the $(2,2)$ mode~\cite{Kapadia:2020kss,Singh:2020lwx}. We aim to explore this point for next-generation detectors such as the Einstein Telescope~\cite{Punturo:2010zz,Branchesi:2023mws,ET:2025xjr} and Cosmic Explorer~\cite{Reitze:2019iox,Evans:2021gyd,Evans:2023euw}, whose low-frequency sensitivity could enable alerts tens of minutes before merger. A natural calibration extension is a P-P-plot test using injections with random noise realizations; a fully realistic version requires running them through a discrete NSBH template bank with a tailored mode-amplitude-ratio library, to capture the suboptimal templates that may be recovered in practice, see Supplemental Material.
Precession is another relevant effect for NSBH binaries and can also help break the $d_L$--$\iota$ degeneracy~\cite{Dhurkunde:2022aek,Fairhurst:2023idl,Zhou:2026hcw}. Incorporating precessing harmonics into our mode-by-mode filtering procedure represents a natural extension of this work.

\acknowledgments

We thank Mukesh Kumar Singh, Hsin-Yu Chen, Sylvia Biscoveanu, Steve Fairhurst, Daniel Holz, Matias Zaldarriaga, Daryl Haggard, Alberto Colombo, and Kara Merfeld for helpful discussions. We also thank Juan Calderón Bustillo for comments on the manuscript.
This research has made use of data, software and/or web tools obtained from the Gravitational Wave Open Science Center (\url{https://www.gw-openscience.org/}), a service of LIGO Laboratory, the LIGO Scientific Collaboration and the Virgo Collaboration. LIGO Laboratory and Advanced LIGO are funded by the United States National Science Foundation (NSF) as well as the Science and Technology Facilities Council (STFC) of the United Kingdom, the Max-Planck-Society (MPS), and the State of Niedersachsen/Germany for support of the construction of Advanced LIGO and construction and operation of the GEO600 detector. Additional support for Advanced LIGO was provided by the Australian Research Council. Virgo is funded, through the European Gravitational Observatory (EGO), by the French Centre National de Recherche Scientifique (CNRS), the Italian Istituto Nazionale di Fisica Nucleare (INFN) and the Dutch Nikhef, with contributions by institutions from Belgium, Germany, Greece, Hungary, Ireland, Japan, Monaco, Poland, Portugal, Spain. The work of F.I. is supported by a Miller Postdoctoral Fellowship. F.I. and E.B. are supported by NSF Grants No.~AST-2307146, No.~PHY-2513337, No.~PHY-090003, and No.~PHY-20043, by NASA Grant No.~21-ATP21-0010, by John Templeton Foundation Grant No.~62840, by the Simons Foundation [MPS-SIP-00001698, E.B.], by the Simons Foundation International [SFI-MPS-BH-00012593-02], and by Italian Ministry of Foreign Affairs and International Cooperation Grant No.~PGR01167.
J.R. acknowledges support from the Jonathan M.\ Nelson Center for Collaborative Research.
This work was carried out at the Advanced Research Computing at Hopkins (ARCH) core facility (\url{https://www.arch.jhu.edu/}), which is supported by the NSF Grant No. OAC-1920103. 

\appendix

\section{Coherent likelihood with higher-order modes}
\label{app:coherent_score}

Here we summarize how the mode-by-mode SNR time series are combined in the marginalized likelihood used in the main text, %
following the formalism of Ref.~\cite{Wadekar:2024zdq}. For a fixed $(2,2)$ template, the detector-frame waveform in detector $k$ can be written as
\begin{widetext}
\begin{equation}
\begin{split}
  h_k(f) \simeq \frac{d_{L,0}}{d_L}
  \left[F_{+,k}\frac{1+\cos^2\iota}{2} - i \cos\iota\,F_{\times,k}\right]
  \bigg[
  e^{2i\phi_0}\mathbb{h}_{22}(f)
  + e^{3i\phi_0}R_{33}\sin\iota\,\mathbb{h}_{33}(f)
  + e^{4i\phi_0}R_{44}\sin^2\iota\,\mathbb{h}_{44}(f)
  \bigg]\;,
\end{split}
\label{eq:hm_predicted_waveform}
\end{equation}
\end{widetext}
where $\mathbb{h}_{\ell m}$ are unit-normalized complex templates for the modes, $d_{L,0}$ is a reference distance for which the $(2,2)$ mode has unit SNR (assuming $F_+=1, F_\times=0, \iota=0$), $d_{L}$ is the luminosity distance, and $F_{+,k}$ and $F_{\times,k}$ are the antenna responses. The ratios $R_{\ell m}$ are the intrinsic mode-amplitude ratios defined in Eq.~\eqref{eq:Rlm}; in practice they are drawn from the precomputed library described in the following. %
We restrict the analysis to $\ell=|m|$ modes, for which the polarization and inclination dependence can be factorized as in Eq.~\eqref{eq:hm_predicted_waveform}.

The public example notebook in our GitHub repository\footnote{\url{https://github.com/JayWadekar/flywheel}.} follows this notation closely. The file \texttt{snrs\_timeseries\_*.hdf5} stores, for each event, the complex matched-filter time series $\rho_{\ell\ell,k}(t)$ for the $(2,2)$, $(3,3)$, and $(4,4)$ templates in each detector, together with the common time grid. The file \texttt{snrs\_opt\_info\_*.hdf5} stores the optimal-SNR normalization for each mode and detector, as well as the normalized cross-mode overlaps $\langle\mathbb{h}_{22}|\mathbb{h}_{33}\rangle$, $\langle\mathbb{h}_{22}|\mathbb{h}_{44}\rangle$, and $\langle\mathbb{h}_{33}|\mathbb{h}_{44}\rangle$. These two files provide the data-dependent term and the template covariance entering the likelihood in Eq.~\eqref{eq:likelihood_definition}. The separate \texttt{sampled*.h5}, \texttt{snrs\_22\_*.txt}, \texttt{snrs\_33\_*.txt}, and \texttt{snrs\_44\_*.txt} files provide the intrinsic samples and the corresponding $R_{33}$ and $R_{44}$ values used for the discrete marginalization over mode ratios (see Sec.~\ref{sec:Mode_Amp_ratio_samples}).

The covariance between the mode templates enters the normalization term in the likelihood. For a single detector,
\begin{widetext}
\begin{equation}
    \langle h|h\rangle
    =\sum_{\ell,\ell'}
    \!\frac{d_{L,0}^2}{d_{L}^2}\!
    \left|F_+\frac{1+\cos^2\iota}{2}\!-iF_\times\cos\iota\right|^2
    \!\!\!e^{i\phi_0(\ell'-\ell)}
  C_{\ell,\ell'}\;,\quad
C\!=\!
\begin{bmatrix}
1 &
\!\!R_{33}\sin\iota\,\langle\mathbb{h}_{22}|\mathbb{h}_{33}\rangle &
\!\!R_{44}\sin^2\iota\,\langle\mathbb{h}_{22}|\mathbb{h}_{44}\rangle\\
- &
\!\!R_{33}^2\sin^2\iota &
\!\!R_{33}R_{44}\sin^3\iota\,\langle\mathbb{h}_{33}|\mathbb{h}_{44}\rangle\\
- &\!\! - &
\!\!R_{44}^2\sin^4\iota
\end{bmatrix}\!,
\label{eq:hm_hh}
\end{equation}
\end{widetext}
where the omitted lower-triangular entries are fixed by Hermitian conjugation.

In the code, these quantities are passed to \texttt{cogwheel} as two arrays. The array \texttt{dh\_mtd} contains the rescaled matched-filter products $(d|h_{\ell m})$ with dimensions [mode, time, detector], ordered as $(22,33,44)$. The array \texttt{hh\_md} contains the six independent covariance entries with dimensions [mode-pair, detector], ordered as $(22,22)$, $(22,33)$, $(22,44)$, $(33,33)$, $(33,44)$, and $(44,44)$. Since the SNR time series are generated from normalized templates, while the \texttt{cogwheel} distance lookup table assumes inner products at its reference distance, the notebook multiplies these arrays by the factor \texttt{dist\_factor\_ref} (or its square for \texttt{hh\_md}). This is a change of normalization convention only; the final luminosity-distance samples are converted back to physical (Mpc) units in postprocessing.

The data-dependent term is evaluated from the complex matched-filter time series for each mode. If $\rho_{\ell\ell,k}(t)$ denotes the inner product of the data in detector $k$ with the unit template $\mathbb{h}_{\ell\ell}$ at time $t$, then
\begin{widetext}
\begin{equation}
\!\!\langle h_k | d_k e^{i2\pi f t}\rangle
\!=\frac{d_{L,0}}{d_{L}}
\!\left[F_{+,k}\frac{1+\cos^2\iota}{2}-i\cos\iota\,F_{\times,k}\right]
\!\!\bigg[
e^{2i\phi_0}\rho_{22,k}(t)
 + e^{3i\phi_0}R_{33}\sin\!\iota\,\rho_{33,k}(t)
 + e^{4i\phi_0}R_{44}\sin^2\!\iota\,\rho_{44,k}(t)
\bigg]\,.
\label{eq:hm_dh}
\end{equation}
\end{widetext}
The arrival time in each detector is $t_k=t_\oplus+\bm{r}_k\cdot\hat{\bm{n}}/c$, where $t_\oplus$ is the geocentric arrival time, $\bm{r}_k$ is the detector position, and $\hat{\bm{n}}$ is the sky direction.

For computational efficiency, the notebook also constructs an incoherent arrival-time proposal from the single-detector SNR time series. This proposal whitens the three mode time series using the cross-mode covariance matrix and forms an approximate single-detector log likelihood over arrival time. It is used only to propose promising detector arrival times before enforcing the coherent sky-location delays in the full likelihood.

For a fixed set of extrinsic parameters and mode ratios, the detector log likelihood is
\begin{equation}
\ln\mathcal{L}_k =
\mathrm{Re}\langle h_k|d_k\rangle
-\frac{1}{2}\langle h_k|h_k\rangle \;.
\label{eq:likelihood_definition}
\end{equation}
The coherent score is obtained by marginalizing the product of detector likelihoods over distance, phase, sky location, inclination, polarization, geocentric time, and the discrete library of mode-amplitude-ratio samples:
\begin{widetext}
\begin{equation}
e^{\rho_{\rm coh}^2/2}
\simeq
\sum_i w^{(i)}
\int d\Pi\,
\exp\left[
\sum_{k\in{\rm dets}}
\ln\mathcal{L}_k
\left(t_k,R_{33}^{(i)},R_{44}^{(i)},\iota,\phi_0,d_{L},\psi,\hat{\bm{n}}\right)
\right]\;.
\label{eq:hm_multidet_coherent_score}
\end{equation}
\end{widetext}

Note that this is the same coherent score used in the ranking statistic of the \texttt{IAS-HM} search pipeline and computed for all the zero-lag and time-slided candidates~\cite{Wadekar:2024zdq}.  
Our likelihood has the same structure for nuisance parameters such as luminosity distance and reference phase: when their dependence is smooth or can be tabulated accurately, integrating them out is faster than sampling them explicitly while preserving the posterior for the parameters retained in the analysis.

\begin{algorithm}[tb]
\small
\caption{Low-latency marginalization with mode-by-mode filtering}
\label{alg:low_latency_hm_pe}
\DontPrintSemicolon
\SetKwInOut{Input}{Input}
\SetKwInOut{Output}{Output}
\Input{$\rho_{\ell\ell,k}(t)$, mode covariances, samples of $(R_{33},R_{44})$, detector network.}
\Output{Posterior samples for sky position, distance, inclination, polarization, time, and intrinsic quantities.}
\BlankLine
\textbf{Precompute:}\;
Rescale $\rho_{\ell\ell,k}(t)$ and covariances to the reference-distance convention.\;
Build an incoherent arrival-time proposal from the whitened mode SNR time series.\;
\BlankLine
\textbf{Marginalize:}\;
\ForEach{QMC point and mode-ratio sample}{
Draw $\iota$, $\psi$, $t_\oplus$, sky position, and $(R_{33},R_{44})$.\;
Map $t_\oplus$ to detector arrival times $t_k$.\;
Interpolate $\rho_{\ell\ell,k}(t)$ at each $t_k$.\;
Evaluate $\langle h_k|d_k\rangle$ and $\langle h_k|h_k\rangle$ using Eqs.~\eqref{eq:hm_hh} and~\eqref{eq:hm_dh}.\;
Marginalize over $d_{L}$ and $\phi_0$.\;
}
\BlankLine
\textbf{Sample:}\;
Combine detectors and weighted mode-ratio samples to estimate the marginalized integral Eq.~\eqref{eq:hm_multidet_coherent_score}.\;
Importance-resample the marginalization points.\;
Repeat with $R_{33}=R_{44}=0$ for the quadrupole-only comparison.\;
\end{algorithm}

In Eq.~\eqref{eq:hm_multidet_coherent_score}, $w^{(i)}$ are the weights associated with the precomputed mode-ratio samples. In the implementation used here, the distance integral is evaluated using a precomputed interpolation table, the $\phi_0$ integral by trapezoid quadrature, and the remaining extrinsic-parameter integrals by adaptive importance sampling. Note that the effect of $\phi_0$ is different for different higher modes, and hence we cannot marginalize the orbital phase
analytically as typically done for quadrupole-only waveforms~\cite{Singer:2015ema}.
Instead, we use trapezoid quadrature, which performs
adequately since the likelihood is a periodic function of
the orbital phase. Once we perform the marginalized integral in Eq.~\eqref{eq:hm_multidet_coherent_score}, we use it to draw the low-latency posterior samples discussed in the main text.

The example notebook evaluates Eq.~\eqref{eq:hm_multidet_coherent_score} twice for the same event. The higher-mode run uses the sampled $R_{33}$ and $R_{44}$ values. The comparison ``$(2,2)$-only'' run sets these ratios to zero, leaving the same detector network, time series, and sampling settings but removing the higher-mode contribution from the waveform model. Posterior samples are then drawn by importance resampling from the quasi-Monte Carlo points used in the marginalization; App.~\ref{sec:marginalization_sample_example} gives a simpler one-template version of the same logic for reference. The notebook reports the sky position, luminosity distance, polarization, inclination, source-frame secondary mass, effective spin, and likelihood. The source-frame mass and effective-spin samples are attached by carrying along the same intrinsic-sample index used for the mode-ratio sample; the luminosity distance is converted back from the \texttt{cogwheel} normalization, and the right ascension is obtained from the sampled longitude and geocentric time.

For the timing estimate quoted in the main text, we selected ten representative O5 catalog events and timed only the marginalization call and the subsequent posterior-sample generation, excluding input loading and construction of the coherent-score object (performed once for the entire O5 run). The benchmark used $512$ phase points, a minimum target effective sample size of $50$, and a maximum of $2^{16}$ QMC samples. The higher-mode runs used $\simeq5.3\times10^3$ QMC samples on average, corresponding to $\simeq2.6\times10^6$ vectorized QMC-phase likelihood evaluations. After the max-over-distance thresholding step, $\simeq2.2\times10^5$ QMC-phase points per event were retained for distance-marginalized likelihood lookup, and the average effective sample size was $\simeq1.9\times10^2$.

\section{Avenues for future work}
\label{app:future_work}

We outlined a few directions for future work at the end of the main text. Here we discuss additional limitations of the present implementation.

First, some high-SNR events can yield a low effective sample size in the adaptive marginalization. This happens because the posterior support becomes narrow enough that the present QMC resolution is not always sufficient; increasing the maximum QMC resolution or adapting the proposal more aggressively should improve the sampling efficiency
for such events.
Second, the current framework assumes that the intrinsic parameters are fixed to the maximum-likelihood template provided by the search pipeline, with the precomputed mode-ratio samples representing the neighborhood of that template. A more complete low-latency treatment could combine information from multiple nearby templates, thereby approximately marginalizing over intrinsic parameter space rather than conditioning on a single best-fit template. This could follow template-bank or iterative-grid approaches used in dot-PE~\cite{Mushkin:2025yks} and RIFT~\cite{Lange:2018pyp,Yelikar:2023mwg,Wofford:2023iwz}.

Third, in our simulations we compute the SNR time series using the quadrupole-mode template evaluated at the injected parameters. A realistic search uses a discrete NSBH template bank, so the recovered trigger may correspond to a slightly mismatched, suboptimal template. This idealization also limits the calibration tests we can currently perform. A standard P-P plot would draw injections from the prior, add a random noise realization to each, and check whether the resulting credible levels are uniformly distributed; with our zero-noise, perfect-template setup neither ingredient is present, and in particular the maximum-likelihood point lies at the injection by construction, biasing any credible-level CDF away from the diagonal. Performing a meaningful P-P test therefore requires two coupled upgrades: adding noise realizations drawn from the assumed O5 PSDs, and recovering each injection through a discrete NSBH template bank so that the suboptimal-template effect is included. The second step also requires a mode-amplitude-ratio library tailored to that bank, constructed using the procedure of App.~\ref{sec:Mode_Amp_ratio_samples} applied to each template rather than only to the best-fit point. We plan to address both upgrades in a follow-up project, either using our \texttt{IAS-HM} search pipeline or in collaboration with one of
the LVK pipelines.

Finally, we included only the dominant $\ell=|m|=3$ and $4$ modes in this analysis. Other HMs can be incorporated in the same framework,
and we leave this extension to future work.

\section{Illustrative example: posterior sampling via marginalization}
\label{sec:marginalization_sample_example}

The advantage of marginalization over brute-force sampling can be illustrated with a simple matched-filtering example. Consider a single-template likelihood with unknown amplitude $A$ and reference phase $\phi_0$,
\begin{equation}
    \ln {\cal L}(A, \phi_0) = A\,{\rm Re}\!\left[e^{-i\phi_0}\langle d|h\rangle\right] - \frac{1}{2}A^2 \langle h|h\rangle\;,
\end{equation}
where $\langle d|h\rangle$ is the (complex) matched-filter inner product of the data with a unit-normalized template. Sampling jointly in $(A, \phi_0)$ requires exploring a banana-shaped posterior, and the $\phi_0$ direction is periodic and can be multimodal. Marginalizing analytically over $\phi_0$ with a uniform prior yields the well-known closed form
\begin{equation}
    {\cal L}_{\rm marg}(A) \propto I_0\!\left(A\,|\langle d|h\rangle|\right) \exp\!\left[-\frac{1}{2}A^2 \langle h|h\rangle\right]\;,
\end{equation}
where $I_0$ is the modified Bessel function of the first kind. This removes a parameter without approximation, eliminates the multimodality induced by $\phi_0$, and leaves a one-dimensional likelihood that can be evaluated directly. This is precisely the structure exploited by \texttt{BAYESTAR} and \texttt{cogwheel} for $\phi_0$, and analogously by a precomputed lookup table for the luminosity distance $d_{L}$. Posterior samples in $(A, \phi_0)$ are then obtained in two steps,
mirroring the procedure used in our pipeline. First, $A_j$ is drawn
from the one-dimensional marginalized posterior
$\mathcal{L}_{\rm marg}(A)\,p(A)$, for example by inverse-CDF sampling on a grid. Second, the conditional posterior for the phase given
$A_j$ is
\begin{equation}
    p(\phi_0\,|\,A_j, d) \propto
    \exp\!\left[A_j|\langle d|h\rangle|\cos(\phi_0 - \phi_d)\right]\;,
\end{equation}
where $\phi_d \equiv \arg\langle d|h\rangle$ is a von Mises
distribution from which $\phi_{0,j}$ can be drawn directly. The pair
$(A_j, \phi_{0,j})$ is then an unweighted joint posterior sample. Our pipeline extends the same logic to the discrete library sampled from Eq.~\eqref{eq:hm_multidet_coherent_score}, replacing stochastic exploration over a hard combinatorial space with a deterministic weighted sum. Note that the HM case is different from this simple example.
The orbital phase $\phi_0$ enters
each harmonic with a different prefactor and cannot be marginalized
analytically as in the $(2,2)$-only
case~\cite{Singer:2015ema}; we instead integrate it numerically by
trapezoid quadrature (see App.~\ref{app:coherent_score}). The toy
example above captures the structure of the resulting two-step
sampling procedure, with the analytic von Mises step replaced by an
inverse-CDF draw from the tabulated phase posterior.

\section{Creating samples for mode amplitude ratios}
\label{sec:Mode_Amp_ratio_samples}

In this appendix, we describe how we construct the library of $R_{\ell m}$ samples from Eq.~\eqref{eq:Rlm} for a given $(2,2)$ template: 
\begin{equation}
R_{\ell m} \equiv \frac{\langle h_{\ell m}(f)|h_{\ell m}(f)\rangle^{1/2}}{\langle h_{22}(f)|h_{22}(f)\rangle^{1/2}}\;.
\label{eq:Rlm_appendix}
\end{equation}
Each $(2,2)$ template in a search pipeline corresponds not to an infinitesimal point, but to a region of intrinsic parameter space [$m_1, m_2, \chi_{1z}, \chi_{2z}$]. We therefore need a fast way to assign $R_{\ell m}$ values to that region. One approach is to perform Monte Carlo sampling over [$m_1, m_2, \chi_{1z}, \chi_{2z}$], collect the samples associated with each template, and calculate $R_{33}$ and $R_{44}$ for those samples. Instead of storing a discrete list of $\{R_{33}, R_{44}\}$ for each template, one can model an interpolated probability density $p(R_{33}, R_{44}|t_\alpha)$, where $t_\alpha$ denotes the parameters of the $(2,2)$ template (e.g., SVD coordinates in the case of Refs.~\cite{Wadekar:2023gea,Roulet:2019hzy}). This normalizing-flow approach was used in the IAS-HM O3 search~\cite{Wadekar:2023gea} and in Ref.~\cite{Zhou:2026hcw} for precession harmonics. The model $p(R_{33}, R_{44}|t_\alpha)$ can be learned during template-bank construction. In this paper, however, we did not build a full NSBH template bank, and instead use the approximate method described below.

For each $(2,2)$ template, we construct a scaled Fisher ellipse to approximate its associated region of intrinsic parameter space.
Consider the match between neighboring points given by $M \equiv \langle h(\theta) | h(\theta +\Delta \theta) \rangle/(||h(\theta)||\ ||h(\theta+\Delta \theta)||) \equiv 1 - g_{ij} \Delta \theta^i \Delta \theta^j$, where $g_{ij}$ is the Riemannian metric in the intrinsic parameters. This metric can be derived from the Fisher information matrix as $g_{ij} = \Gamma_{ij}/(2\rho^2)$, where $\Gamma_{ij} = \langle \partial_i h| \partial_j h \rangle$ and $\rho \equiv \langle h | h \rangle^{1/2}$ is the SNR. The region of parameter space corresponding to each template is given by the mismatch ellipse~\cite{Owen:1998dk}
\begin{equation}
g_{ij}\Delta \theta^i \Delta \theta^j <  \mu \;,  
\end{equation}
where $\mu$ is the threshold mismatch. We use $\mu=1-0.97=0.03$, corresponding to the minimal match commonly adopted in template-bank construction. We then sample intrinsic parameters inside the mismatch ellipse and construct a library of $\{R_{33}, R_{44}\}$ values for use in the marginalized integral in Eq.~\eqref{eq:hm_multidet_coherent_score}. 

A useful byproduct of this construction is that each
$(R_{33}^{(i)}, R_{44}^{(i)})$ sample is generated from a specific
intrinsic-parameter point
$\theta^{(i)} = (m_1^{(i)}, m_2^{(i)}, \chi_{1z}^{(i)}, \chi_{2z}^{(i)})$
inside the mismatch ellipse. Storing $\theta^{(i)}$ alongside the
mode ratios lets us obtain approximate posteriors on derived
intrinsic quantities, such as the secondary mass $m_2^{(i)}$, chirp
mass $\mathcal{M}^{(i)}$, mass ratio $q^{(i)}$, and effective spin
$\chi_{\rm eff}^{(i)}$, at essentially no additional cost. We obtain
importance weights on $(R_{33}^{(i)}, R_{44}^{(i)})$ from
Eq.~\eqref{eq:hm_multidet_coherent_score} and use the same weights on
the associated $\theta^{(i)}$, so the intrinsic posteriors follow by
reweighting the stored samples. Because the $\theta^{(i)}$ are drawn
from inside the Fisher mismatch ellipse, the implicit prior is uniform within that neighborhood of the best-fit template
(an astrophysical prior can be applied as a postprocessing
reweighting if desired). This is also the origin of the mild
$m_2^{\rm src}$ overconstraint visible in Fig.~\ref{fig:GW190814}:
the intrinsic posterior is conditioned on the best-fit-template
neighborhood, not marginalized over the full intrinsic parameter
space (see App.~\ref{app:future_work}).

\section{Simulated population and uncertainties}\label{app:population}

\begin{figure}[tb]
    \centering
    \includegraphics[width=\linewidth]{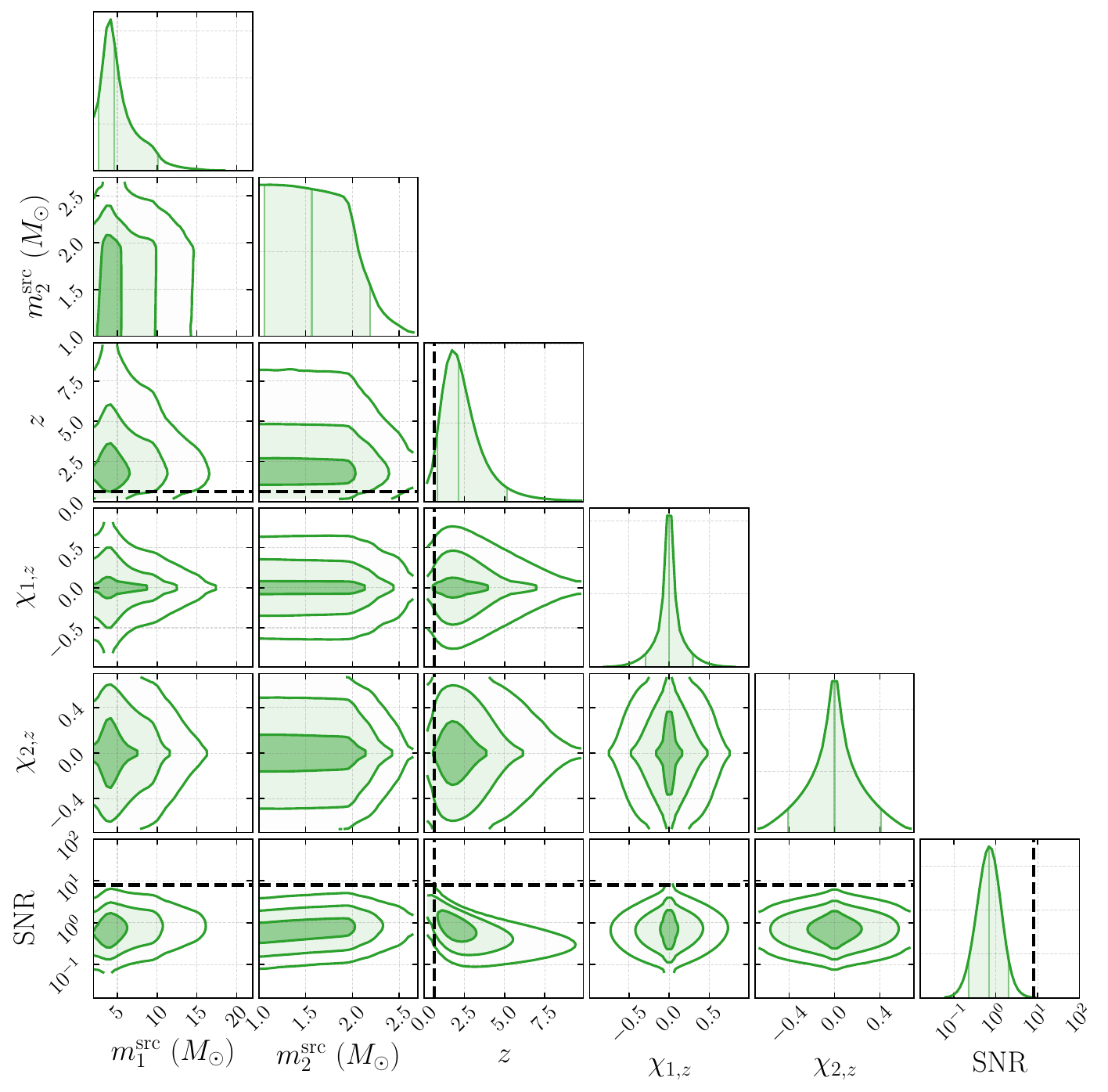}
    \caption{In the main text, we tested our low-latency HM pipeline on mergers detected in 100 simulated one-year NSBH catalogs with O5 LIGO--Virgo design-sensitivity network. Here, we show the intrinsic parameters, redshift, and SNR of all the events (detected+undetected) in the combined set of our catalogs; dashed lines mark the furthest detected-event redshift and the SNR threshold.}
    \label{fig:corner_allnsbhPop}
\end{figure}

We simulate 100 realizations of a catalog corresponding to 1\,yr of observation time. The source-frame masses $m_{1,2}^{\rm src}$ and spins are drawn according to the \textsc{NSBH-pop} model first introduced in Ref.~\cite{Biscoveanu:2022iue}. In this model, the BH distribution is assumed to be a truncated power-law, while the mass ratio distribution is assumed to be a truncated Gaussian between $1\,{\rm M}_\odot/m_1^{\rm src}<q<{\rm min}(m_{\rm NS,\,max}/m_1^{\rm src}, 1)$, where $m_{\rm NS,\,max}$ denotes the maximum NS mass. The BH spin magnitudes are modeled as a Beta distribution in the range $\chi_1\in[0,\,0.99]$, while the NS spin magnitudes are assumed to be uniformly distributed over $\chi_2\in[0,\,0.7]$. The cosines of the spin tilts are sampled uniformly.
Each one of our realizations corresponds to a different (hyper)posterior sample from the latest LVK analysis~\cite{LIGOScientific:2024elc}. From the same sample we also extract the local merger rate and low-$z$ slope of the merger rate distribution. We extend the distribution to high redshift using a Madau-Dickinson profile~\cite{Madau:2014bja}, such that 
\begin{equation}
    p(z) \propto \dfrac{{\rm d}V_{\rm c}}{{\rm d}z}(z) \dfrac{(1+z)^{\alpha_z-1}}{1 + \left(\dfrac{1+z}{1+z_p}\right)^{\alpha_z+\beta_z}}\;.
\end{equation}
We use the typical parameters $z_p=2$ and $\beta_z=3$~\cite{Madau:2016jbv}. The redshift of the furthest detection is $z\sim0.64$, so this high-redshift extension does not affect our results. Figure~\ref{fig:corner_allnsbhPop} shows the resulting intrinsic-parameter distribution across the 100 catalogs, along with the SNR in the chosen LVK design-sensitivity network. The extrinsic parameters are sampled uniformly in their physical ranges.

\begin{figure}[tbp]
    \centering
    \includegraphics[width=\linewidth]{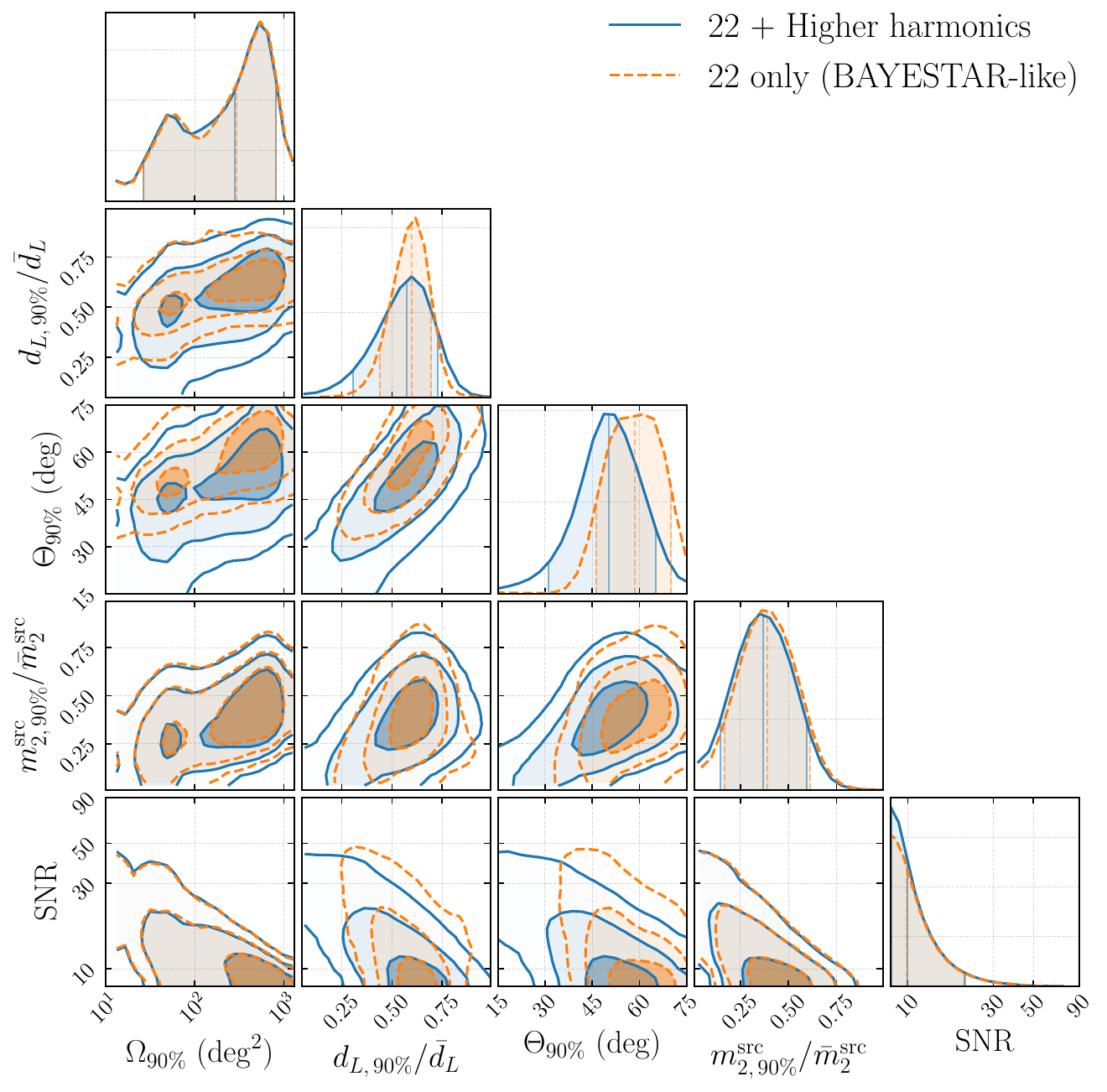}
    \caption{Population-level one-dimensional summaries complement the event-by-event ratios shown in the main text. We compare the HM pipeline (blue) with the $(2,2)$-only pipeline (orange dashed) for sky area, relative $d_L$ and $m_2^{\rm src}$ widths, viewing-angle width, and SNR. We see that HMs mainly improve distance and viewing-angle information, while sky area and SNR remain largely unchanged.}
    \label{fig:corner_errors_distributions_comparison}
\end{figure}

The main text focuses on relative improvements between the HM and $(2,2)$-only pipelines. As a complement, Fig.~\ref{fig:corner_errors_distributions_comparison} reports the distributions of the 90\% localization area, the relative 90\% widths of the luminosity-distance and source-frame-secondary-mass posteriors, the 90\% width of the viewing-angle posterior, and the optimal matched-filter SNR. Sky localization and SNR are largely unchanged by HMs, while the improvement in $m_2^{\rm src}$ is mild, as discussed in the main text. The viewing-angle distribution shifts appreciably to smaller values with the full pipeline, and the $d_L$ distribution also gains support at smaller distances. For a minority of systems, the relative width of the $d_L$ posterior is larger in the full analysis. This occurs partly because the HM posterior can shift to smaller $d_L$ (see, e.g., Fig.~\ref{fig:GW190814}), giving a larger relative uncertainty even when the absolute posterior width is comparable to the $(2,2)$-only result. In some low-SNR systems, HMs can also make the posterior structure slightly more complex, producing the mild broadening visible in the $d_{L,\,90\%}/\bar{d}_{L}$--SNR panel of Fig.~\ref{fig:corner_errors_distributions_comparison}.

\section{Example of an event with larger \texorpdfstring{$V_{\rm c}$}{Vc} when including higher-order modes}\label{sec:largerVc22example}

\begin{figure}[tbp]
    \includegraphics[width=\linewidth]{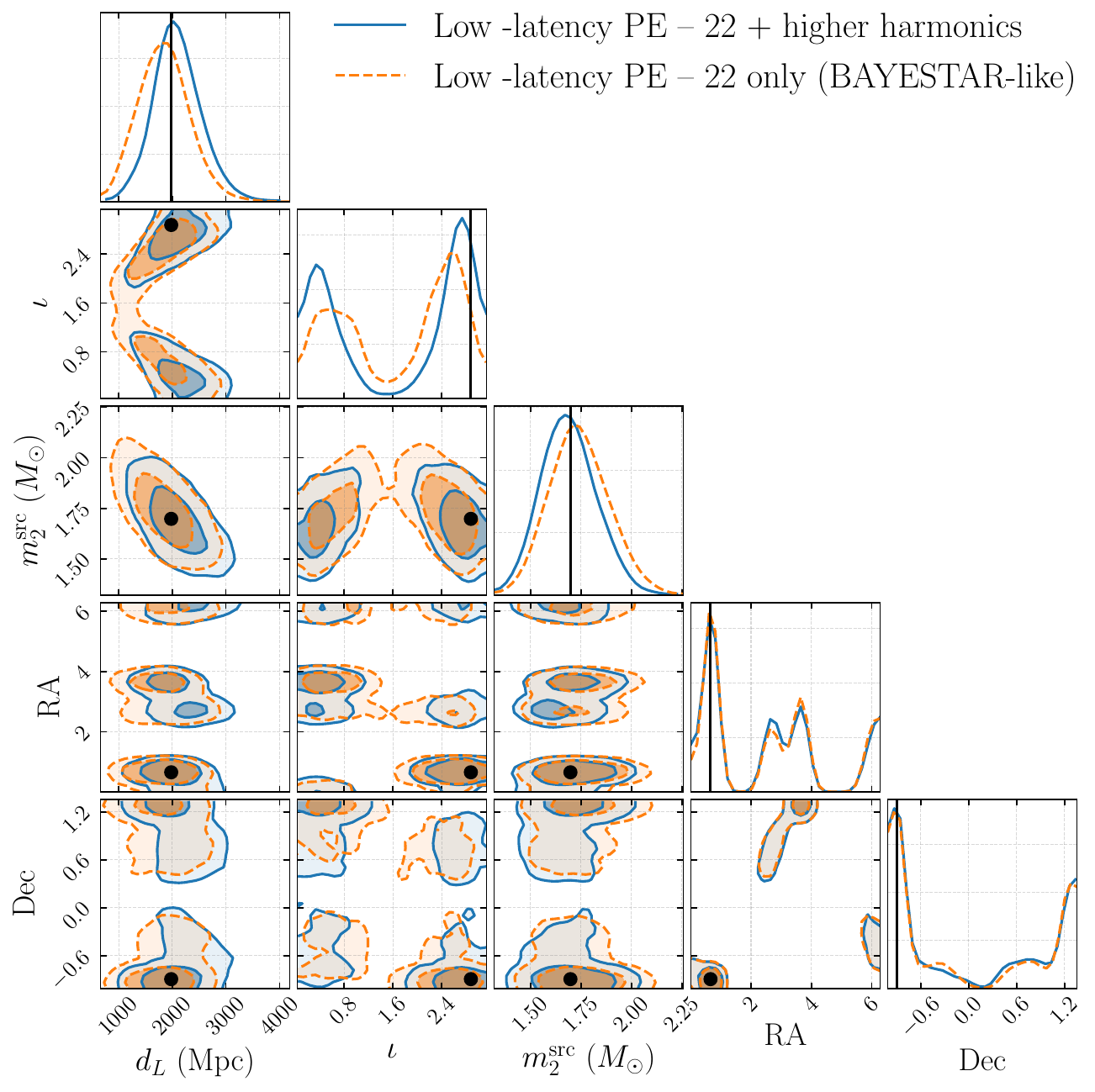}
    \caption{A small fraction of events in the main text figure about the $d_L$--$\iota$ constraints showed a larger 90\% c.l. localization comoving volume when HMs are included. Here, we show one example of such events. We see that the HM posterior is closer to the injected luminosity distance. Therefore, a larger quoted volume is not necessarily a failure of the HM analysis, because the posterior can become less biased while shifting to distances with larger comoving volume.}
    \label{fig:corner_example_event_betterVc22}
\end{figure}

As reported in the main text, for a small fraction of the simulated events we find a larger localization comoving volume at 90\% c.l. when running the analysis including HMs. This is not a spurious effect: for systems with small viewing angles and little observed harmonic content, accounting for subdominant harmonics reduces support for edge-on configurations. The posterior can therefore shift toward larger luminosity distances, where the corresponding comoving volume is larger. While the 90\% $V_{\rm c}$ is larger using the full pipeline, the posterior peaks closer to the true value. Figure~\ref{fig:corner_example_event_betterVc22} shows one such example, in which the relative 90\% $d_L$ posterior from the full analysis is $\sim\!8\%$ narrower, while the 90\% localization comoving volume is $\sim\!12\%$ larger. The sky-localization posteriors are consistent between the two analyses, but the full analysis gives more support to the injected parameters.

\clearpage

\bibliographystyle{apsrev4-1-etal}
\bibliography{Main}

\begin{thebibliography}{113}
\bibitem [{zer()}]{zeronoise_explain}%
  \BibitemOpen
  \href@noop {} {}\bibinfo {note} {In a zero-noise injection, the SNR of an
  injected signal is defined relative to the detector's assumed noise power
  spectral density, even though no random noise realization is added. It
  measures the signal's strength compared to the noise level, not noise
  actually present in the data.}\BibitemShut {Stop}%
\bibitem [{pep()}]{pepriors_explain}%
  \BibitemOpen
  \href@noop {} {}\bibinfo {note} {The standard choices in GW PE are a prior
  uniform in $\cos\iota$, and a prior uniform in comoving volume $V_{\rm c}$
  and source-frame time for $d_L$. The former assigns more prior mass to
  edge-on configurations when expressed as a density in $\iota$, while the
  latter strongly favors large distances. These two effects partially compete:
  edge-on systems are intrinsically fainter and therefore, for a given observed
  amplitude, tend to be inferred at smaller distances, whereas the volumetric
  distance prior pushes the posterior toward larger $d_L$. In general, if the
  likelihood does not constrain the inclination well, the distance posterior
  can be significantly influenced by these priors, often being shifted toward
  larger distances by the $d_L$ prior.}\BibitemShut {Stop}%
\end{thebibliography}
\end{document}